\pdfoutput=1
\documentclass[a4paper,unpublished,noshowkeys,nopdfoutputerror]{quantumarticle}
\usepackage{amsmath,amssymb,bm}
\usepackage{graphicx}
\usepackage{microtype}
\usepackage{hyperref}

\begin{document}

\title{A Gauge Sign Rule for Quantum Rotor Networks}
\author{Swagata Acharya}
\affiliation{National Laboratory of the Rockies, Golden, CO 80401, USA}

\begin{abstract}
The sign problem is the non-positive path-integral weight that makes a quantum
system hard to simulate classically. It shows up in problems that look
unrelated, from fermion antisymmetry and geometric frustration to real-time
evolution, topological angles, and finite density. We study it here for compact
continuous variables, modeled as networks of quantum rotors,
$\hat H=\sum_i 4E_C(\hat n_i-n_{g,i})^2-\sum_{\langle ij\rangle}
E_{ij}\cos(\hat\phi_i-\hat\phi_j-\theta_{ij})$. For these, a single
gauge-invariant quantity controls the obstruction, the frustration flux through
the independent loops of the interaction graph. We prove a sign rule, the rotor
analogue of Marshall's. The charge-basis Hamiltonian is sign-free if and only if
every loop flux vanishes modulo $2\pi$. Exact diagonalization and density-matrix
renormalization then show that the sign cost is a gauge-invariant function of
the flux. For a single loop it vanishes at zero flux, peaks at $\pi$, and is
exponentially small in the loop's perimeter. Summed over many loops it is
extensive, growing with system size. The same loop flux generates the sign of three problems
usually treated apart, namely the Mott transition (through the slave-rotor
mapping), compact $U(1)$ lattice gauge theory, and frustrated continuous
optimization. A superconducting rotor array realizes both the model and its sign
natively.
\end{abstract}
\maketitle

\section{Introduction}

The recurring signature of quantum hardness is not a system but a mathematical
object, a path integral whose measure is not real and positive,
$Z=\int\mathcal D\phi\,e^{-S[\phi]}$ with $e^{-S}\notin\mathbb R_{\ge0}$. The
same non-positive measure arises from fermion antisymmetry, geometric
frustration, real-time evolution, topological $\theta$-terms and Berry phases,
and finite density. A classical stochastic simulation needs non-negative
weights, so a non-positive measure forces it to pay an exponential price, the
sign problem~\cite{TroyerWiese2005,Loh1990,BlankenbeclerScalapinoSugar1981}. A
quantum processor, which propagates complex amplitudes, does not. A companion
paper~\cite{AcharyaPRXQ} used this idea to build an analog rotor solver for one
specific problem. Here we address the general question behind it. For continuous
systems, we ask where the sign obstruction lives and how to compute it.

For a network of quantum rotors, the sign is controlled entirely by the
frustration flux through the loops of the interaction graph, a single
gauge-invariant quantity. The charge-basis Hamiltonian is free of the sign
problem exactly when every loop flux vanishes. This is the continuous analogue
of Marshall's rule for antiferromagnets. When the flux is nonzero, the sign has
a cost. That cost is a gauge-invariant function of the flux. It is exponentially
small for a single frustrated loop, but extensive when many loops are frustrated.
The same loop flux generates the sign of three problems usually treated apart, namely
the Mott metal--insulator transition, compact lattice gauge theory, and frustrated
optimization. A superconducting rotor array realizes the model, and its sign,
natively.

\section{Model and the sign rule}

\subsection{Rotor networks and the loop flux}

Consider quantum rotors on a graph $\mathcal G=(\mathcal V,\mathcal E)$, each
with a compact phase $\hat\phi_i\in[-\pi,\pi]$ conjugate to an integer charge
$\hat n_i$, $[\hat\phi_i,\hat n_j]=i\delta_{ij}$,
\begin{equation}
\hat H=\sum_{i\in\mathcal V} 4E_C(\hat n_i-n_{g,i})^2
-\sum_{\langle ij\rangle\in\mathcal E} E_{ij}\cos(\hat\phi_i-\hat\phi_j-\theta_{ij}),
\label{eq:rotor}
\end{equation}
with $E_{ij}>0$. The link phases $\theta_{ij}=-\theta_{ji}$ form a compact $U(1)$ connection on
the graph, physically a magnetic flux, a spin--orbit phase, or, in the optimization
setting, loop congestion. In the charge eigenbasis $\{|\{n_i\}\rangle\}$ the
charging term is diagonal and the Josephson term moves one Cooper pair across a
link. Writing $\cos(\hat\phi_i-\hat\phi_j-\theta_{ij})=\tfrac12
(e^{-i\theta_{ij}}e^{i(\hat\phi_i-\hat\phi_j)}+\mathrm{h.c.})$ and using
$e^{i\hat\phi_i}|n_i\rangle=|n_i+1\rangle$,
\begin{equation}
\begin{split}
&\langle \dots,n_i+1,n_j-1,\dots|\,\hat H\,|\dots,n_i,n_j,\dots\rangle\\
&\quad=-\tfrac12 E_{ij}\,e^{-i\theta_{ij}}.
\end{split}
\label{eq:hop}
\end{equation}
A sign-free (stoquastic) representation needs this element to be real and
non-positive~\cite{BravyiTerhal2008}. The only obstruction to that is the phase
$e^{-i\theta_{ij}}$. But the individual $\theta_{ij}$ are
gauge-dependent. The basis rotation $|n_i\rangle\to e^{i\chi_i n_i}|n_i\rangle$,
or equivalently $\hat\phi_i\to\hat\phi_i+\chi_i$, shifts
$\theta_{ij}\to\theta_{ij}+\chi_i-\chi_j$ and leaves Eq.~\eqref{eq:rotor}
invariant. What survives this freedom are the loop fluxes,
\begin{equation}
W_\mathcal C=\sum_{(ij)\in\mathcal C}\theta_{ij}\pmod{2\pi},
\end{equation}
one for each independent cycle $\mathcal C$ of $\mathcal G$. These are the
complete gauge invariants of the connection, and the quantities that fix the
sign.

\subsection{The sign rule}

The sign structure of Eq.~\eqref{eq:rotor} in the charge basis is fixed by the
loop fluxes. The rule is simple. The charge basis is sign-free if and only if
every independent loop flux vanishes, $W_\mathcal C\equiv0\pmod{2\pi}$. Here
sign-free means that a phase rotation of the basis states can bring every
off-diagonal element to real, non-positive form. This is the continuous-variable,
compact-$U(1)$ counterpart of Marshall's rule for bipartite
antiferromagnets~\cite{Marshall1955}. Marshall removes the sign of an
unfrustrated antiferromagnet by a sublattice rotation. Here the absence of
frustration is the absence of loop flux, and a rotation of the same kind does
the job. When the flux does not vanish, no such rotation exists. The argument
holds on any graph and for arbitrary diagonal changes of phase, not only linear
gauges, and it follows the motion of a single Cooper pair.

In the charge basis the charging term is diagonal, so the only off-diagonal
matrix elements are the single-pair hops of Eq.~\eqref{eq:hop},
\[
\langle \mathbf n'|\hat H|\mathbf n\rangle=-\tfrac12 E_{ij}\,e^{-i\theta_{ij}},
\qquad \mathbf n'=\mathbf n+\mathbf e_i-\mathbf e_j,
\]
one for each edge, together with the reverse hop. Because $E_{ij}>0$, such an
element is real and non-positive exactly when the link phase $\theta_{ij}$ is a
multiple of $2\pi$. We are free to redefine the phase of each basis state,
$|\mathbf n\rangle\to e^{if(\mathbf n)}|\mathbf n\rangle$, with any real function
$f$. These diagonal rotations keep the charge labels intact, so they are the
only changes of basis for which stoquasticity in this basis is even meaningful.
Such a rotation leaves the charging term alone and multiplies each hop by a
phase. The Hamiltonian is then sign-free exactly when $f$ can be chosen to
absorb every link phase,
\begin{equation}
f(\mathbf n')-f(\mathbf n)\equiv\theta_{ij}\pmod{2\pi},
\label{eq:cocycle}
\end{equation}
edge by edge.

Whether that is possible is decided by the loops. Carry a Cooper pair once
around a closed cycle $\mathcal C$ and it returns to the configuration it started
from. Add Eq.~\eqref{eq:cocycle} around the loop. The left-hand side cancels
term by term, while the right-hand side adds up to the loop flux $W_\mathcal C$.
A loop that carries flux therefore cannot be cured on all its edges at once, and
no rotation of the basis, however nonlinear, can remove the sign. If instead
every loop flux vanishes, the link phase has no circulation. A circulation-free
field is a gradient (the discrete Poincar\'e lemma), so there is a site potential
$\alpha_i$ with $\theta_{ij}=\alpha_i-\alpha_j$. Assign $\alpha$ freely along a
spanning tree and propagate it. The remaining edges close consistently precisely
because the fluxes vanish. The linear choice $f(\mathbf n)=\sum_i\alpha_i n_i$
then cancels every link phase and leaves the Hamiltonian sign-free. The gate
charges $n_{g,i}$ sit on the diagonal throughout and play no part.

Two consequences follow. On a tree, with no loops, the model is
always sign-free, because the sign problem is a property of cycles. And within diagonal
unitaries the rule is exact. There the loop flux is a genuine obstruction,
not merely a sufficient one. Whether a wholly different representation, such as a
nonlocal change of basis or a dual model, could remove a given flux is the hard
open question. Deciding it in general is expected to be intractable, since curing the
sign problem is NP-complete in the worst
case~\cite{MarvianLidarHen2019,Hangleiter2020}, though special structures admit
exact cures~\cite{ChandrasekharanWiese1999}. We therefore call frustration
\emph{removable} when an efficient transformation undoes it, and \emph{protected}
when it is topologically irreducible. The natural measure of severity is the rate
at which the average sign of the natural stochastic estimator decays,
$\langle s\rangle\sim e^{-N\Delta(W)}$.

\section{The sign rate}

\subsection{A gauge-invariant function of the flux}

\begin{figure}[t]
\centering
\includegraphics[width=\linewidth]{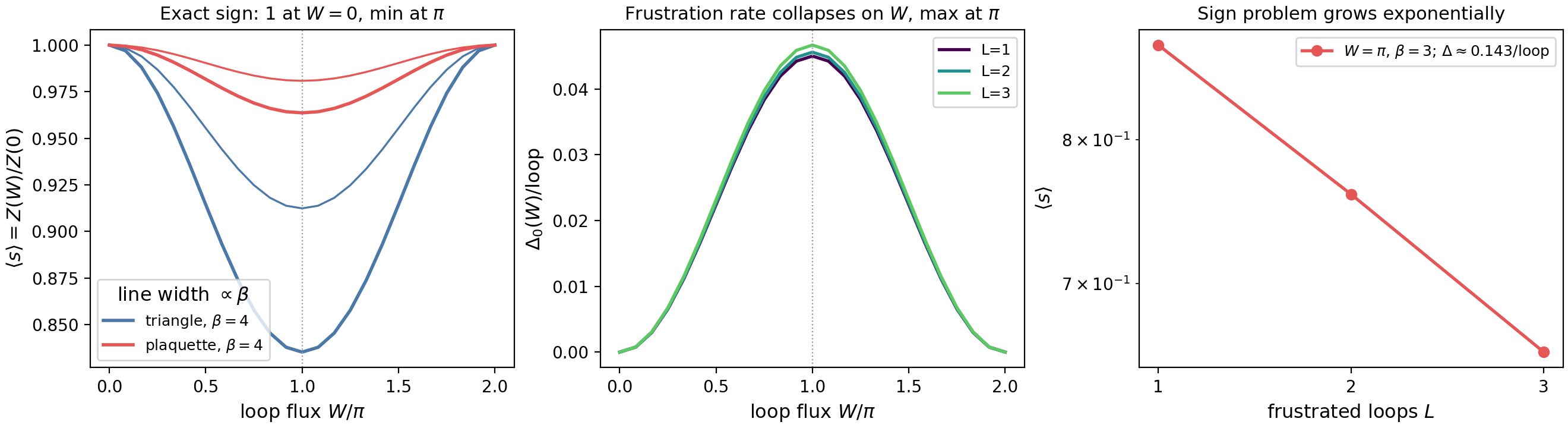}
\caption{The sign is a gauge-invariant function of the loop flux (exact
diagonalization). \textit{Left:} the finite-$T$ average sign
$\langle s\rangle=Z(W)/Z(0)$ for a triangle and a plaquette, unity at $W=0$, a
minimum at $W=\pi$, deeper at larger $\beta$ (line width $\propto\beta$).
\textit{Centre:} the frustration rate $\Delta_0(W)/\mathrm{loop}$ for $L=1,2,3$
frustrated loops collapses onto a single function of the flux, zero at $W=0$ and
maximal at $W=\pi$. \textit{Right:} at $W=\pi$ the average sign decays
exponentially with the number of loops, $\langle s\rangle\sim e^{-\Delta L}$.}
\label{fig:first}
\end{figure}

We test the rule and its quantitative refinement by exact diagonalization of
Eq.~\eqref{eq:rotor} in the charge basis, with each rotor truncated to five
charge states ($|n_i|\le2$), using a matrix-free ground-state
solver. Three features emerge (Fig.~\ref{fig:first}). The exact finite-temperature average sign
$\langle s\rangle=Z(W)/Z(0)$ is unity at zero flux, never exceeds it, and falls
to a minimum at $W=\pi$ that deepens as the temperature drops. The
zero-temperature frustration rate is $\Delta_0(W)=E_0(W)-E_0(0)$. It controls
the low-temperature sign through $\langle s\rangle\sim e^{-\beta N\Delta_0}$,
which we verify directly ($-\log\langle s\rangle/\beta N\to\Delta_0/N$). This
rate is a \emph{single} function of the loop flux. The curves for one, two, and
three frustrated loops collapse onto one another, vanish at $W=0$, and peak at
$W=\pi$. And at $\pi$ flux the average sign falls exponentially with the number
of loops, a genuine, size-growing sign problem.

Gauge invariance is exact, as the rule requires, and the numerics bear it out
to machine precision. Placing a fixed loop flux on a single edge, or spreading it
over three, gives the same $\langle s\rangle$. On two different two-loop
lattices, an arbitrary gauge transformation
$\theta_{ij}\to\theta_{ij}+\alpha_i-\alpha_j$ scrambles the individual link
phases while preserving the loop fluxes. It leaves the ground-state energy
unchanged for every random choice of the site potentials $\alpha_i$. Sweeping the two independent fluxes of such a
lattice maps a smooth surface $\Delta_0(W_1,W_2)$ (Fig.~\ref{fig:origin}, right)
that vanishes only on the trivial locus $W_1=W_2=0$ and rises monotonically to
its maximum at $(\pi,\pi)$. The sign rate is thus a function of the loop fluxes
alone, positive off the trivial locus and greatest at $\pi$.

\subsection{Perimeter law and extensive rate}

\begin{figure}[t]
\centering
\includegraphics[width=\linewidth]{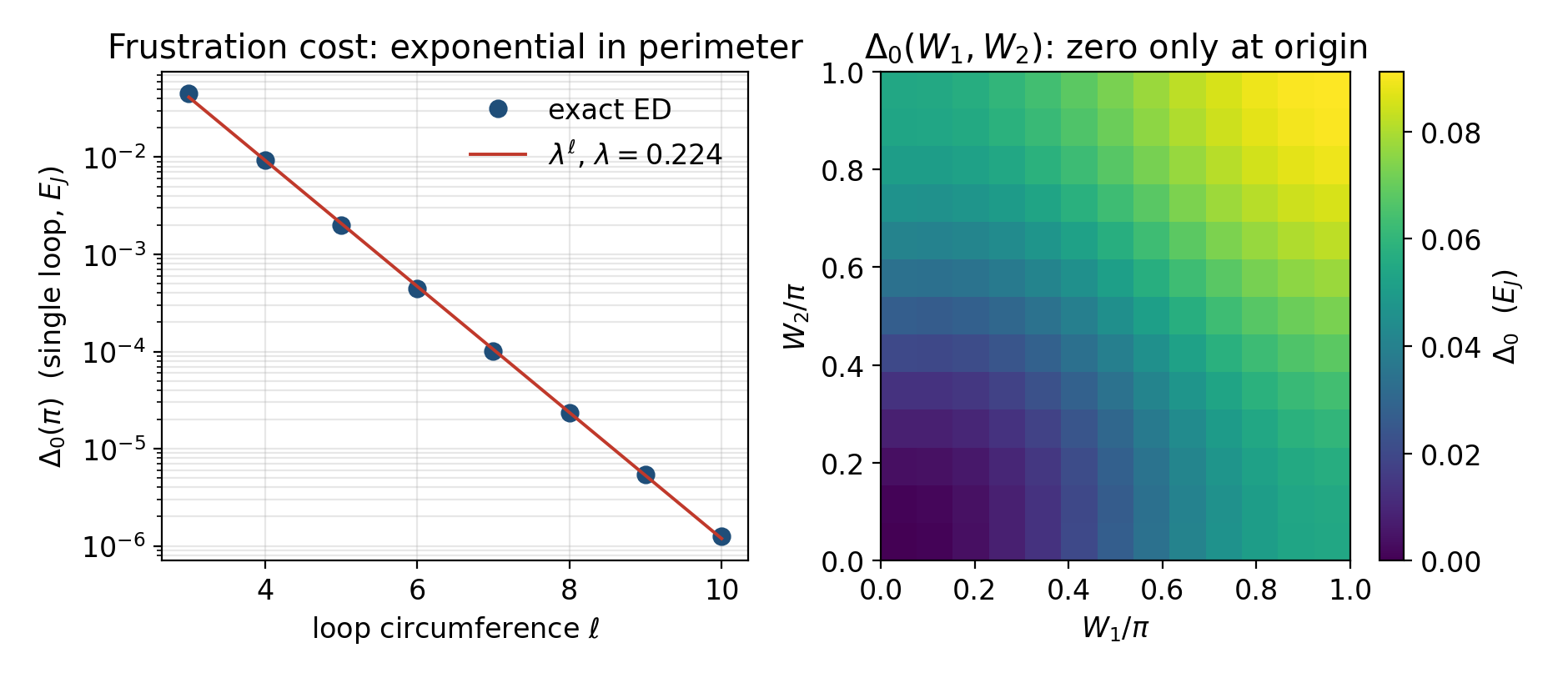}
\caption{Microscopic origin and exact gauge invariance. \textit{Left:} the cost
of a single $\pi$-flux loop decays exponentially with its circumference $\ell$,
$\Delta_0(\pi)\sim\lambda^\ell$, $\lambda\approx0.22$ (semilog, line is the
fit), an around-the-loop instanton of order $\ell$ in the hopping.
\textit{Right:} on a two-loop lattice the rate $\Delta_0(W_1,W_2)$ is a smooth
function of the two loop fluxes alone, zero only at the origin and maximal at
$(\pi,\pi)$, and invariant under any gauge gradient on the links.}
\label{fig:origin}
\end{figure}

The magnitude of the sign rate has a microscopic origin. Take a
single frustrated loop and vary its circumference $\ell$. The cost
$\Delta_0(\pi)$ of one $\pi$-flux loop falls exponentially with the perimeter,
$\Delta_0(\pi)\sim\lambda^{\ell}$ with $\lambda\approx0.22$. On a log plot this
is a straight line from $\ell=3$ to $\ell=10$, more than four decades in
$\Delta_0$ (Fig.~\ref{fig:origin}, left). Physically, to register the flux at
all, charge must virtually circulate once around the loop. This is a process of
order $\ell$ in the hopping, an around-the-loop instanton whose amplitude is
exponentially small in the loop length. The shortest loops
dominate, which is why a triangle ($\ell=3$) is the most sign-problematic
elementary frustration and a square ($\ell=4$) already about five times
milder.

\begin{figure}[t]
\centering
\includegraphics[width=\linewidth]{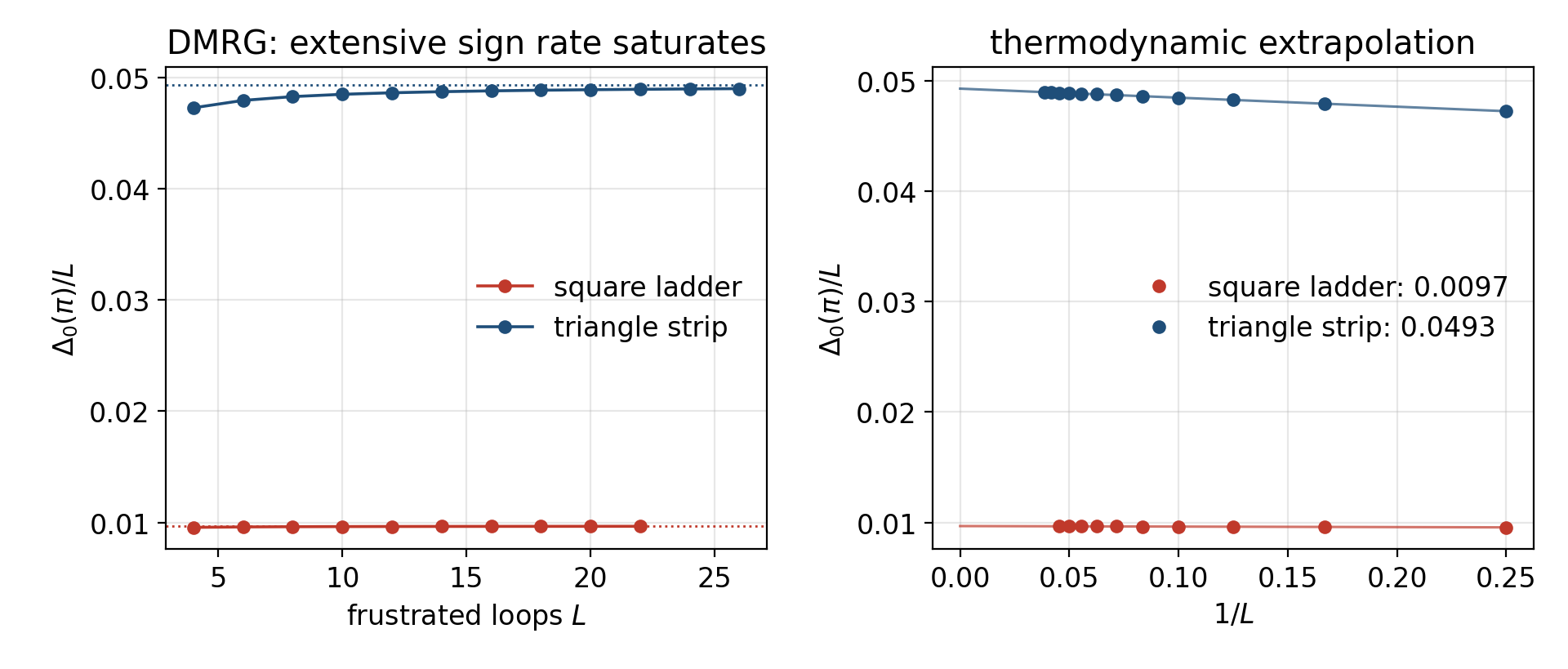}
\caption{The extensive rate, measured past the exact-diagonalization wall with
DMRG. \textit{Left:} the per-loop rate $\Delta_0(\pi)/L$ saturates with $L$ for
both ladders, reaching the thermodynamic value directly (the square ladder
attains $0.00965$ by $L=20$, $N=42$). \textit{Right:} the $1/L$ view with the
linear extrapolation. The DMRG intercepts agree with the exact-diagonalization
estimates. Our ground-state exact diagonalization reaches $N\simeq11$
($\sim5\times10^7$ states). Its cost grows as $5^N$, so it stays far short of the
$N=42$ ($5^{42}$) ladders reached by DMRG.}
\label{fig:dmrg}
\end{figure}

Do many such loops add up to an extensive sign problem, or wash out? This is
settled by going to large systems. We use two extensively frustrated geometries,
a triangular strip (every triangle frustrated) and a square ladder (every
plaquette frustrated). In both, the per-loop rate $\Delta_0(\pi)/L$
\emph{saturates} as the number of loops $L$ grows. The sign problem is therefore
extensive, with a finite rate per loop. Exact diagonalization reaches $N=11$
sites. Density-matrix renormalization (DMRG) pushes the same ladders far past
that wall and confirms the saturation directly. The square ladder reaches
$\Delta_0(\pi)/\mathrm{loop}=0.00965$ already at $L=20$ ($N=42$, a Hilbert space
of $5^{42}$ states). The thermodynamic rate is therefore measured, not merely
extrapolated (Fig.~\ref{fig:dmrg}). The two rates are
$\Delta_0(\pi)/\mathrm{loop}\approx0.049$ (triangular) and $0.0097$ (square), in
units of the Josephson coupling. Both are geometry-dependent but strictly
positive, and their ratio follows the perimeter law, since the single-loop values
reproduce the $L=1$ end of each ladder exactly. A finite, lattice-specific sign
rate turns on the moment the loops are frustrated, and vanishes only on the
gauge-trivial locus.

\subsection{Removable versus protected frustration}

The distinction between removable and protected frustration is, concretely, a
count of saddle points. The collective coordinate of a single frustrated
loop is the phase circulating around it, a particle on a ring threaded by the
Aharonov--Bohm flux $W$. Its ground state feels the flux only through
trajectories that wind once around the ring, the same around-the-loop instanton
responsible for the perimeter law. There are two such saddles, winding $+1$ and
$-1$. They carry opposite Aharonov--Bohm phases $\pm W$ and interfere as
$\cos W$, so the leading flux dependence of the rate is
\begin{equation}
\frac{\Delta_0(W)}{\Delta_0(\pi)}=\frac{1-\cos W}{2}+\mathcal O(\lambda^{2\ell}),
\end{equation}
destructive and maximal at $W=\pi$. In the language of Lefschetz
thimbles~\cite{Witten2011,Cristoforetti2012,Alexandru2022}, at $W=0$ a single real thimble
dominates and the weight is positive. The sign is removable, and the
Hamiltonian is sign-free by the rule above. For $W\neq0$ two thimbles of equal
magnitude contribute with a relative phase set by $W$, and their interference is
a genuine, irreducible sign. This is protected frustration, greatest at $\pi$. A
stochastic or complex-Langevin estimator fails precisely when the second thimble
acquires comparable weight, and the phase that makes the interference
irreducible is exactly the gauge-invariant loop flux.

\begin{figure}[t]
\centering
\includegraphics[width=\linewidth]{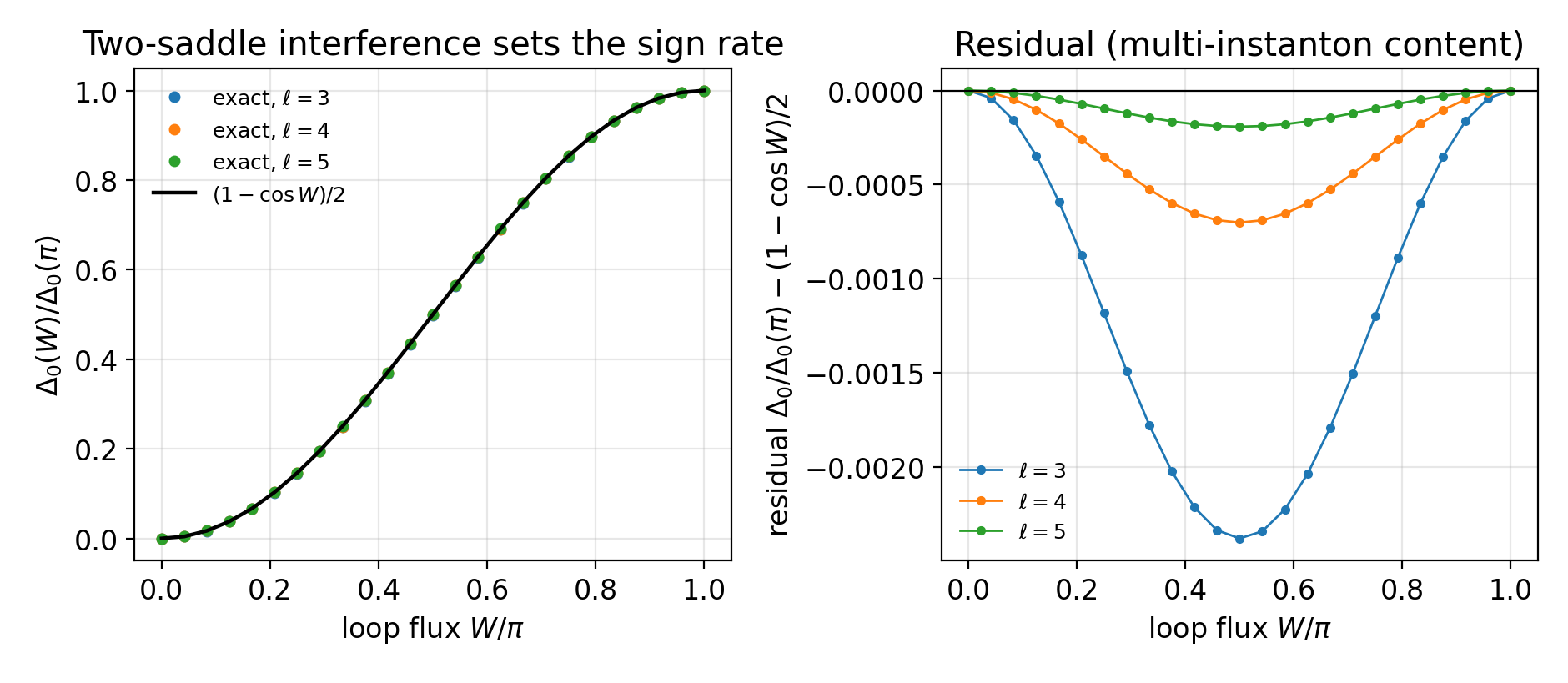}
\caption{Removable versus protected frustration (single loop, exact
diagonalization). \textit{Left:} $\Delta_0(W)$ for circumferences $\ell=3,4,5$
collapses onto the two-saddle interference form $(1-\cos W)/2$ (black), zero at
$W=0$ (one thimble, removable) and maximal at $W=\pi$ (two thimbles in antiphase,
protected). \textit{Right:} the residual is a $\cos 2W$ two-instanton
correction, suppressed by a further $\lambda^\ell$ and hence shrinking with loop
size.}
\label{fig:thimble}
\end{figure}

Exact diagonalization confirms this (Fig.~\ref{fig:thimble}). The rate
$\Delta_0(W)$ for loop circumferences $\ell=3,4,5$ collapses onto
$(1-\cos W)/2$ to better than $0.3\%$, and the small residual is a
second-harmonic $\cos 2W$ two-instanton term that shrinks with loop size, as the
perimeter law demands, since each additional winding costs a further factor
$\lambda^\ell$.

\section{One invariant, three faces}

The same loop flux is the sign generator of three problems usually studied
apart.

\paragraph{Hubbard/Mott via slave rotors.} In the slave-rotor
representation~\cite{FlorensGeorges2004,KotliarRuckenstein1986,Acharya2016} the electron is
fractionalized as $c_{i\sigma}=e^{-i\theta_i}f_{i\sigma}$, with a compact charge
rotor $e^{i\theta_i}$ and a spinon $f_{i\sigma}$, subject to
$\hat L_i=\sum_\sigma f^\dagger_{i\sigma}f_{i\sigma}-\tfrac12$. The hopping
becomes $c^\dagger_i c_j=f^\dagger_i f_j\,e^{i(\theta_i-\theta_j)}$. Decoupling
the spinon sector yields exactly Eq.~\eqref{eq:rotor} for the charge rotors,
with $E_C\sim U$ and $E_{ij}\sim t_{ij}\langle f^\dagger_i f_j\rangle$. The
Mott transition is the rotor superfluid--insulator transition. The quasiparticle
weight is the rotor condensate fraction $|\langle e^{i\theta}\rangle|^2$, nonzero
in the metal and zero in the Mott insulator. But the intersite coupling $E_{ij}$,
set by the spinon bond order, persists into the insulator, so the rotor
Hamiltonian and its loop fluxes stay nontrivial on both sides. Frustration of the lattice, orbital loops,
or spin--orbit coupling induces a nonzero rotor Wilson loop. The Mott problem
then inherits our flux invariant, and the sign-problematic regime of slave-rotor
quantum Monte Carlo is exactly where $W_\mathcal C\neq0$. The split into rotor
and spinon is not unique. It carries a $U(1)$ gauge freedom
($\theta_i\to\theta_i+\alpha_i$, $f_{i\sigma}\to e^{i\alpha_i}f_{i\sigma}$) that
generates an emergent gauge field $a_{ij}$ on the links, so the rotor hopping
appears as $\cos(\theta_i-\theta_j-a_{ij})$
\cite{LeeNagaosaWen2006,Senthil2008}. The loop sum of this field,
$\oint_\mathcal C a=W_\mathcal C$, is again the loop flux, so the Mott problem
and lattice gauge theory are one Hamiltonian in two guises. Because the sign
depends only on that gauge-invariant flux, and never on how the electron was
split, the obstruction is physical, not an artifact of the slave-particle
construction.

\paragraph{Compact gauge theory and the $\theta$-term.} Placing rotors on the
links of a lattice gives compact $U(1)$ lattice gauge
theory~\cite{Polyakov1977}. The loop flux $W_\mathcal C$ is the plaquette Berry
phase, and a topological $\theta$-term or real-time evolution makes that weight
complex. These are the archetypal sign problems of high-energy physics
($\theta$-vacua, finite density, real-time gauge dynamics), and the rotor rule
identifies them as the same obstruction on link variables.

\paragraph{Frustrated continuous optimization.} The frustrated $XY$/rotor-glass
partition function~\cite{KosterlitzThouless1973} (power-grid state estimation in Ref.~\cite{AcharyaPRXQ}, but
equally orientational glasses and synchronization) is Eq.~\eqref{eq:rotor} in
its classical limit, with loop congestion in the role of $W_\mathcal C$. Here
the sign rate $\Delta(W)$ is not a statement about the classical optimization,
which carries a positive Boltzmann weight and is frequently easy, but about the
quantum evaluation of the same partition function, where the loop flux is the
obstruction a sampling method must confront.

\subsection{A worked example (triangular-lattice organics)}

Take the half-filled Hubbard model on the anisotropic triangular lattice, the
accepted minimal model of the organic Mott insulators
$\kappa$-(BEDT-TTF)$_2X$~\cite{PowellMcKenzie2011}. Slave-rotor mean
field~\cite{FlorensGeorges2004} writes $c_{i\sigma}=e^{-i\theta_i}f_{i\sigma}$
and decouples the charge rotor from the spinon. The rotor sector is exactly
Eq.~\eqref{eq:rotor},
\begin{equation}
H_\theta\simeq\frac U2\sum_i\hat L_i^2
-\sum_{\langle ij\rangle} t_{ij}\,|Q^f_{ij}|\,
\cos\!\big(\theta_i-\theta_j-\theta^{\rm eff}_{ij}\big),
\end{equation}
with charging energy $E_C=U/2$, Josephson coupling $E_{ij}=t_{ij}|Q^f_{ij}|$ set
by the spinon bond order $Q^f_{ij}=\langle f^\dagger_{i}f_{j}\rangle$, and link
phase $\theta^{\rm eff}_{ij}=\arg Q^f_{ij}+a_{ij}$, the spinon-bond phase plus
the emergent gauge field of the decomposition.

The sign rule now reads off the charge sector's sign structure directly. The
charge sector is sign-free precisely when the per-triangle flux
$W_\triangle=\sum_{\triangle}\theta^{\rm eff}_{ij}$ vanishes, that is, when the emergent spinon state carries no loop flux. A
time-reversal-symmetric, uniform spin liquid (real $Q^f$, no gauge flux) leaves
$W_\triangle=0$, and the charge sector is sign-free. A \emph{chiral} spin liquid,
or any spontaneous or field-induced flux, switches on $W_\triangle\neq0$ and a
sign rate $\Delta_0(W_\triangle)$ read from
Fig.~\ref{fig:origin}. The material's sign problem is thereby tied to a
measurable property, broken time reversal, i.e.\ scalar spin chirality. Consider
$\kappa$-(BEDT-TTF)$_2$Cu$_2$(CN)$_3$, a candidate gapless spin
liquid~\cite{Shimizu2003}. Its $U/t\approx6$--$8$ places it just inside the Mott
insulator. There $|Q^f|\sim0.1$--$0.2$, giving $E_{ij}/E_C=2t|Q^f|/U\approx
0.05$--$0.1$, deep on the charge-localized side, where the rate is fixed purely
by $W_\triangle$. A chiral instability
(under field or pressure) should make the charge sector's classical simulation
cost jump as $e^{N\Delta_0(W_\triangle)}$, while a genuinely non-chiral spin
liquid should stay sign-free. This is a correspondence between Hamiltonians, not
a solved material. Whether $W_\triangle$ is actually nonzero, and hence whether
the charge sector really has a sign problem, is decided by the true ground state
of the spinons and their gauge field. The mean-field treatment used here does
not determine it.

\section{Discussion}

\subsection{Native quantum realization}

Equation~\eqref{eq:rotor} is directly the Hamiltonian of a superconducting
circuit. A Josephson/fluxonium array~\cite{Manucharyan2009} with
charging energy $E_C$, Josephson couplings $E_{ij}$, and flux-biased loops
setting $\theta_{ij}$ realizes the model with $W_\mathcal C$ programmable in
situ. The device propagates the complex amplitude of Eq.~\eqref{eq:hop} by
construction, evolving the non-positive measure that a classical sampler cannot
follow~\cite{Feynman1982,Lloyd1996}.

\subsection{Sign versus entanglement}

Two different obstructions make a quantum system hard to simulate classically.
The first is the sign problem. Methods that work by sampling, such as quantum
Monte Carlo~\cite{Gull2011}, turn the quantum weight into a probability and
average over random configurations. This only works if the weight stays
positive. When it does not, positive and negative contributions cancel. The
number of samples needed to see through the cancellation then grows
exponentially with size and inverse temperature, exactly the rate $\Delta_0$
computed above. Sampling does
not care about spatial dimension. A sign-free model is tractable in one, two, or
three dimensions alike. For Monte Carlo, the sign, not the dimension, is the
wall, and curing it in general is NP-hard~\cite{TroyerWiese2005,MarvianLidarHen2019}.
The sign problem is specific to sampling. A deterministic optimizer has no sign
problem, but it escapes the symptom rather than the hardness. It returns an
optimum, not the partition function or the real-time evolution. And finding even
the ground state of a general local Hamiltonian is itself intractable (formally
QMA-complete).

The second obstruction is entanglement. Methods that instead store the
wavefunction, the tensor networks (matrix-product states) used above, pay a
cost set by how entangled the state is across a cut. In one dimension a gapped
ground state obeys an area law, so the entanglement across a cut is bounded and the
state compresses efficiently. This is a theorem~\cite{Hastings2007}, and is
why the ladder calculations converge easily. In two or three dimensions
the entanglement across a cut grows with the cut, the representation cost grows
exponentially with the system's width, and contracting the network is provably
hard~\cite{Schuch2007}. This obstruction has nothing to do with signs. A
perfectly sign-free two-dimensional system can still lie beyond wavefunction
methods.

There are thus two axes (a problem can be sign-hard,
entanglement-hard, both, or neither), and our rule speaks only to the first. As
a statement it is dimension-blind. It depends on the loops in the lattice, not on
how the lattice sits in space, though a higher-dimensional lattice offers more
independent loops to frustrate. We study quasi-one-dimensional ladders so
that the entanglement axis is benign (the area law holds) and the sign axis is
isolated and can be measured. A genuinely two- or three-dimensional
frustrated array is hard on both axes at once. That is the regime a native rotor
device is built for. It neither samples (no cancellation) nor stores the
wavefunction (no entanglement cost), but holds the amplitude and lets it evolve.

Two familiar hard problems have the same structure. In dynamical mean-field
theory, the continuous-time Monte Carlo solver~\cite{Gull2011} develops a sign
problem when the hybridization acquires complex, off-diagonal elements, for
example from spin--orbit coupling or multi-orbital and cluster geometries. It
worsens as the temperature falls. In our language those complex hopping phases
are loop fluxes, and the low-temperature decay of the average sign follows the
same $e^{-\beta N\Delta_0}$ law. The piece our bosonic rule does not capture is
the separate sign from fermionic exchange. Real-time evolution is the extreme
case. Continuation to real time makes the weight a pure phase on every element
of spacetime, frustration everywhere at once. That is why classical real-time
simulation is generically the hardest of all. A rotor device, by contrast,
evolves in real time by construction.

\section{Conclusion}

For quantum rotor networks the sign problem is governed by the loop flux. The
charge basis is sign-free exactly when the flux vanishes. When the flux is
nonzero, the sign has a cost set by the flux alone, a gauge invariant. That rate
is exponentially small in a single loop's perimeter. Summed over loops it is
extensive, growing with system size. The same invariant governs three problems usually
treated apart, namely the Mott transition, compact lattice gauge theory, and
frustrated optimization. A superconducting rotor array realizes it natively.

The rule is exact within
charge-basis-preserving transformations. Whether a nonlocal or dual
representation can cure a protected flux, and how the rate $\Delta(W)$ maps onto
the complexity classes of the sign problem, remains open. It also governs only the sign of geometric origin, and is silent about
the sign that comes from Fermi statistics, whose relation to the same invariant
we leave open. The unification through slave rotors is a correspondence between Hamiltonians,
not a solved material. In a real Mott insulator the loop flux is not an input but
is generated self-consistently by the spinons and their gauge field. Our result
gives the sign rate as a function of that flux. Which value a given material
realizes is a separate question. The decisive
experiment is a superconducting rotor array swept in flux across $\Delta(W)=0$ at
fixed accuracy. The classical sampling cost rises as $e^{N\Delta(W)}$, while a
native rotor device does not sample and so does not pay it.

A single gauge-invariant flux thus draws the boundary between the classically
simulable and the classically hard for a class of problems that spans correlated
matter, gauge theory, and optimization.

\section{Numerical methods}

All results use the rotor Hamiltonian Eq.~\eqref{eq:rotor} in the charge basis
with a per-site truncation $|n_i|\le n_{\max}$ (local dimension
$d=2n_{\max}+1$). Finite-temperature average signs are computed from the full
spectrum within each conserved-charge sector. Ground-state frustration rates
$\Delta_0(W)=E_0(W)-E_0(0)$ use a matrix-free Lanczos/ARPACK solver, which
reaches $N\simeq11$ sites (Hilbert-space dimension up to $\sim5\times10^{7}$).
The truncation is converged. $\Delta_0(\pi)$ evaluated at $n_{\max}=2,3,4$ agrees
to a part in $10^{6}$ ($n_{\max}=2$ versus $3$) and $10^{9}$ ($n_{\max}=3$ versus
$4$), and all production runs use $n_{\max}=2$. The three geometries are a single
ring, a triangular strip (each triangle frustrated through its diagonal edge),
and a square ladder (each plaquette frustrated through its top leg).

Beyond the exact-diagonalization wall we use matrix-product-state (DMRG) ground
states~\cite{White1992,Schollwock2011}, with adjacent sites blocked in pairs so that the range-two rotor
couplings become nearest neighbour. An odd site count carries a single decoupled
padding site that contributes nothing to $\Delta_0$. The DMRG energies reproduce
exact diagonalization to better than $10^{-7}$ for both ladders at $L\le4$, and
reach $N=42$ (a Hilbert space of $5^{42}$ states) at bond dimension up to
$\chi=400$. Thermodynamic rates follow from a linear fit in $1/L$, with the fits
on the last three and last four points coinciding to the quoted digits. Gauge
invariance is checked by applying random discrete gradients
$\theta_{ij}\to\theta_{ij}+\alpha_i-\alpha_j$ and confirming that the
ground-state energy is unchanged to machine precision.

\section*{Acknowledgements}

This work was authored by the National Laboratory of the Rockies for the U.S.
Department of Energy (DOE) under Contract No.\ DE-AC36-08GO28308. Funding was
provided by the Office of Science, Basic Energy Sciences, Division of Materials,
U.S. Department of Energy. SA acknowledges the use of the National Energy
Research Scientific Computing Center, under Contract No.\ DE-AC02-05CH11231 using
NERSC award BES-ERCAP0021783, and also acknowledge that a portion of the research
was performed using computational resources sponsored by the Department of
Energy's Office of Energy Efficiency and Renewable Energy and located at the
National Laboratory of the Rockies. The views expressed in the article do not
necessarily represent the views of the DOE or the U.S. Government. The U.S.
Government retains and the publisher, by accepting the article for publication,
acknowledges that the U.S. Government retains a nonexclusive, paid-up,
irrevocable, worldwide license to publish or reproduce the published form of this
work, or allow others to do so, for U.S. Government purposes.

\section*{Data and code availability}

The exact-diagonalization and matrix-product-state codes, and the scripts that
generate all figures, are available from the author on reasonable request.

\end{document}